\documentclass[12pt]{JHEP3}
\usepackage{amsmath}
\usepackage{amssymb}
\usepackage{bm}
\usepackage[dvips]{graphicx}

\title{Hawking Radiation and  Quantum Anomaly in AdS$_2$/CFT$_1$ Correspondence}

\preprint{TIFR/TH/08-45 }

\author{
 Takeshi Morita\\

  \vspace{5mm}

\emph{ Department of Theoretical Physics, Tata Institute of Fundamental Research,} \\
\emph{Homi Bhabha Rd, Mumbai 400005, India}\\ 
   \email{takeshi@theory.tifr.res.in}
 }

\abstract{
In order to understand a boundary description of Hawking radiation in the AdS/CFT correspondence, we investigate the trace anomaly method in AdS$_2$ space.
In this method, Hawking radiation is derived from the trace anomaly of the energy-momentum tensor in the bulk.
We find a correspondence between the energy-momentum tensor and a composite operator in  CFT$_1$ and understand the anomalous properties of the energy-momentum tensor in terms of this composite operator.
By using this correspondence, we reproduce Hawking radiation from the boundary description.
In addition, we find a correspondence between higher-spin currents in the bulk and composite operators in the boundary.}

\begin{document}

\section{Introduction}
\setlength{\baselineskip}{7mm}

\setcounter{footnote}{0}
\setcounter{equation}{0}

Black holes are an important topic in the study of quantum field theory.
By considering quantum effects of matter field in a black hole background, it is argued that black holes evaporate through Hawking radiation \cite{Hawking:1974sw,Hawking:1974rv}.
In case of a complete evaporation, the information in the matter which composes the black hole apparently disappears and it seems that unitarity is broken.
This problem is called the information paradox and it should be resolved by quantum gravity or string theory.
One possible solution for this issue is proposed in the AdS/CFT correspondence in string theory \cite{Maldacena:1997re, Gubser:1998bc, Witten:1998qj, Aharony:1999ti}.
In the AdS/CFT correspondence, the dynamics of supergravity fields in the asymptotically AdS space is described by the corresponding boundary conformal field theory which preserves unitarity.
Therefore we can expect that the black hole evaporation would also be understood as a unitary process in the boundary.
In order to achieve this proposal, it is important to understand how to describe Hawking radiation in terms of the boundary conformal field theory.
The purpose of this letter is to find the boundary description of Hawking radiation via the trace anomaly method which was shown by Christensen and Fulling \cite{Christensen:1977jc}.

Christensen and Fulling calculated the energy flux which black holes radiate by studying anomalous properties of energy-momentum tensor of free massless matter fields.
If we consider a black hole in an asymptotically AdS space, such matter fields in the bulk can be described by the corresponding boundary operators \cite{Gubser:1998bc, Witten:1998qj}.
We will investigate the representation of the matter energy-momentum tensor in terms of the boundary operators by employing the bulk-boundary correlation function correspondence \cite{Banks:1998dd,
Balasubramanian:1999ri,Hamilton:2005ju}, and will express the anomalous properties in the bulk in terms of the boundary operators.

Furthermore, we can reproduce the full spectrum of thermal Hawking radiation by calculating anomalies in higher-spin currents. 
This derivation is a generalization of the trace anomaly method \cite{Iso:2007kt,Iso:2007hd}.
In order to reproduce it, we will discuss the construction of these higher-spin currents from the boundary operators.

In this letter, we study a black hole in 2 dimensions for simplicity and assume the existence of the corresponding one dimensional conformal field theory.
Such duality is called AdS$_2$/CFT$_1$ correspondence \cite{Strominger:1998yg,Spradlin:1999bn} and recent developments are in \cite{Azeyanagi:2007bj,Hartman:2008dq, Sen:2008yk, Gupta:2008ki, Sen:2008vm, Cadoni:2008mw,Alishahiha:2008tv,Castro:2008ms}.
The outline of this letter is as follows.
In section 2, we introduce the two dimensional AdS black hole.
We briefly explain the trace anomaly method in its context and calculate the energy flux from the energy-momentum tensor in the bulk.
In section 3, we show that the bulk energy-momentum tensor can be represented in terms of the corresponding boundary operators and reproduce the calculation of the trace anomaly method in terms of the boundary.
In section 4, we argue the construction of the higher-spin currents from the boundary operators.
Section 5 is discussions and conclusions.
In the appendices, we review some properties of the AdS/CFT correspondence and a massless scalar field in two dimensions.

\section{Trace anomaly method in two dimensional black hole}
\setcounter{equation}{0}

In this letter, we study the following two dimensional metric \cite{Hamilton:2005ju}\footnote{
$AdS_2$ is also obtained in the near-horizon metric of Reissner-Nordstrom black holes (see for example  \cite{Spradlin:1999bn}).
We can apply the argument of this letter to this system also and obtain a similar result.}
\begin{align} 
ds^2=-f(r)dt^2+f(r)^{-1}dr^2,~(r>0),~~f(r)=r^2-r_0^2.
\label{2dbh}
\end{align} 
This metric can be obtained through the dimensional reduction of a BTZ black hole \cite{Banados:1992wn} and it is equivalent to Rindler coordinates on AdS$_2$.
The position of the horizon $r_0$ is related to the BTZ mass as $M=r_0^2/8G_3$ in units where three dimensional cosmological constant is unity and the surface gravity $\kappa$ is given by $\kappa = r_0$.
The tortoise coordinate $r_*$ is defined by
\begin{align} 
r_*\equiv\int \frac{dr}{f(r)}  =\frac{1}{2 r_0}\log   \frac{r-r_0}{r+r_0},
\end{align} 
in the outer region ($r\ge r_0$).
Note that $r_*$ is always negative and behaves $r_* \rightarrow 0$ as $r \rightarrow \infty$.
We define $(u,v)$ coordinates as $u=t-r_*$ and $v=t+r_*$, and the Kruskal coordinates $(U,V)$ as $U=-e^{-\kappa u}$ and $V=e^{\kappa v}$.

Now we derive Hawking radiation through the trace anomaly method \cite{Christensen:1977jc}.
We consider a two-dimensional free massless scalar field in the black hole background (\ref{2dbh}).
We will calculate the vacuum expectation value of the $uu$ component of the matter energy-momentum tensor\footnote{We can define two energy-momentum tensors for a massless field in two dimensional theory. One is called \textit{holomorphic energy-momentum tensor} and another is \textit{covariant energy-momentum tensor} (see \cite{Iso:2007nc} for more details).
In this letter, we mainly consider the holomorphic energy-momentum tensor.
We will discuss the difference between these two tensors in section 5.} $\langle T(u) \rangle $ which represents the out-going energy flux.
Note that our system is not an actual black hole but a Rindler space and the terminology ``Hawking radiation" is not correct in this sense.
However, the flux $\langle T(u) \rangle $ is related to energy flux radiated from the BTZ black hole \cite{Setare:2006hq,Xu:2006tq} and we use this terminology in this letter.

The energy-momentum tensor satisfies the trace anomaly ${T^\mu}_\mu=cR/24\pi$ in general, where $c$ denotes the central charge which is $1$ in our case and $R$ denotes the background curvature.
By using this equation, we can exhibit an anomalous transformation of the holomorphic energy-momentum tensor $T(z)$ under a conformal map $z \mapsto w(z)$:
\begin{align} 
\left( \frac{\partial w}{\partial z}\right)^2 T(w)=T(z)+\frac{1}{24 \pi} \{ w,z \}.
\label{transformation}
\end{align} 
Here $\{w,z\}$ is the Schwarzian derivative,
\begin{align} 
\{w,z\}=\frac{w'''}{w'}-\frac{3}{2}\left( \frac{w''}{w'} \right)^2 .
\end{align} 
Now we apply this anomalous transformation law (\ref{transformation}) to a coordinate transformation from $(u,v)$ to the Kruskal $(U,V)$, and obtain
\begin{align} 
\kappa^2 U^2T(U)= T(u)-\frac{1}{48\pi}\kappa^2 . 
\label{transformation-energy-momentum}
\end{align} 
We consider the vacuum expectation value of this equation.
Then we notice that $\langle T(U) \rangle $ diverges at the horizon $U=0$ unless the right-hand side is zero.
Since Kruskal coordinates have no singularity at the horizon, $\langle T(U) \rangle $ should be finite there.
This regularity of $\langle T(U) \rangle $ determines the value of $\langle T(u) \rangle $ as $\kappa^2/48\pi$ near the horizon.
In addition, the background metric does not depend on the time $t$.
Therefore $\langle T(u) \rangle $ does not depend on the time as well.
It implies
\begin{align} 
\langle T(u) \rangle =\frac{1}{48\pi}\kappa^2,
\label{energy-flux}
\end{align} 
in the outer region of the black hole.
We can show that this flux is consistent with the thermal radiation of the free massless scalar field from a black body with temperature $\kappa/2\pi$ \cite{Setare:2006hq,Xu:2006tq}.
We conclude that the anomalous transformation (\ref{transformation}) and the regularity condition reproduce Hawking radiation.

\section{Boundary description of bulk energy-momentum tensor and Hawking radiation}
\setcounter{equation}{0}

We now discuss boundary description of Hawking radiation via the Lorentzian AdS/CFT correspondence \cite{ Banks:1998dd, Balasubramanian:1998sn, Balasubramanian:1998de}.
We have shown that the energy flux associated with Hawking radiation can be derived from the anomalous transformation of the energy-momentum tensor.
We will derive the boundary description of the energy-momentum tensor and then discuss the interpretation of Hawking radiation in terms of boundary operators.

The holomorphic energy-momentum tensor of a free massless scalar field $\phi$ in the bulk is constructed as
\begin{align} 
T(u)=:\partial_u \phi(u,v) \partial_u \phi(u,v):
\equiv \lim_{\epsilon\rightarrow 0} \left(\partial_u \phi(u+\epsilon/2,v) \partial_u \phi(u-\epsilon/2,v)+\frac{1}{4\pi}\frac{1}{\epsilon^2} \right). 
\label{energy-momentum}
\end{align} 
Here we have employed the point splitting method to regularize the singularity of the correlation function.
We consider the boundary description of this bulk operator.
It is known that bulk correlation functions of free scalar fields are expressed in terms of correlation functions of boundary operators in general \cite{Banks:1998dd,Balasubramanian:1999ri,Hamilton:2005ju}(See also appendix \ref{appendix correlation}).
In our case, the bulk correlation function of $\phi$ can be represented by
\begin{align} 
\langle \phi(u_1,v_1) \phi(u_2,v_2) \rangle=\int dt_1 \int dt_2
K(t_1 | u_1,v_1)K(t_2 | u_2,v_2)\langle \mathcal{O}(t_1) \mathcal{O}(t_2) \rangle.
\label{correlation}
\end{align} 
Here $\mathcal{O}(t)$ is the boundary operator dual to $\phi(u,v)$ in the AdS$_2$/CFT$_1$ correspondence and it has conformal dimension $1$ as in (\ref{dual-operator}).
$K(t'|u,v)$ is the smearing function of the normalizable mode of $\phi$ which is given by 
\begin{align} 
K(t'|u,v)=\frac{1}{2}\left( \theta(t'-u)-\theta(t'-v) \right) .
\label{smearing function}
\end{align} 
See appendix \ref{appendix smearing } for the derivation of this smearing function.

Now we can relate the bulk energy-momentum tensor to the boundary correlation function.
By employing (\ref{correlation}), the vacuum expectation of (\ref{energy-momentum}) is represented by 
\begin{align}
\langle T(u) \rangle=& \lim_{\epsilon\rightarrow 0}\left[ \int d\tau_1 d\tau_2
\partial_u K(\tau_1 | u+\epsilon/2,v) \partial_u K(\tau_2 | u-\epsilon/2,v)\langle \mathcal{O}(\tau_1) \mathcal{O}(\tau_2) \rangle+\frac{1}{4\pi}\frac{1}{\epsilon^2} \right] \nonumber \\
=& \frac{1}{4}\lim_{\epsilon\rightarrow 0}\left[ \langle \mathcal{O}(u+\epsilon/2) \mathcal{O}(u-\epsilon/2)\rangle +\frac{1}{\pi}\frac{1}{\epsilon^2}\right] .
\label{energy-momentum-boundary-0}
\end{align}
In this equation, since the singular part of the correlator $ \langle \mathcal{O}(u+\epsilon/2)\mathcal{O}(u-\epsilon/2) \rangle$ behaves $-1/\pi\epsilon^2$ \cite{Hamilton:2005ju, Spradlin:1999bn}, the right-hand side of the last equation can be regarded as an expectation value of a regularized composite operator in the boundary and we define it as $\langle:\mathcal{OO}(u):\rangle$.
Then this equation becomes simply
\begin{align} 
 \langle T(u) \rangle=\frac{1}{4}\langle:\mathcal{OO}(u):\rangle
 \label{energy-momentum-boundary}
 \end{align} 
and it implies that the bulk energy-momentum tensor $\langle T(u) \rangle$ at $(u,v)$ (or equivalently $(t,r_*)$) is related to the boundary composite operator $\frac{1}{4}\langle:\mathcal{OO}(u):\rangle$ at $u=t-r_*$.

According to the correspondence (\ref{energy-momentum-boundary}), the derivation of the Hawking radiation in the previous section should be reproduced in terms of the boundary.
Now we confirm it explicitly.
First we consider a coordinate transformation $u \mapsto U(u)$.
Then the operator $\mathcal{O}$ transforms as $\mathcal{O}^{(u)}(u)=\partial_u U \mathcal{O}^{(U)}(U)$, since it has conformal dimension $1$, and we can rewrite (\ref{energy-momentum-boundary-0}) as
\begin{align}
&\langle:\mathcal{OO}(u):\rangle \nonumber \\
=&\lim_{\epsilon\rightarrow 0} \Biggl[\partial_u U(u+\epsilon/2) \partial_u U(u-\epsilon/2) \ \nonumber \\
&\times \left( 
 \langle \mathcal{O}^{(U)}(U(u+\epsilon/2)) \mathcal{O}^{(U)}(U(u-\epsilon/2))\rangle +\frac{1}{\pi}\frac{1}{(U(u+\epsilon/2)-U(u-\epsilon/2))^2}
 \right)  \nonumber \\
 &-\frac{1}{\pi}\frac{\partial_u U(u+\epsilon/2) \partial_u U(u-\epsilon/2)}{(U(u+\epsilon/2)-U(u-\epsilon/2))^2}  +\frac{1}{\pi}\frac{1}{\epsilon^2} \Biggr] \nonumber \\
 =&\left(\partial_u U \right)^2 \langle:\mathcal{O}^{(U)}\mathcal{O}^{(U)}(U)  :\rangle
- \lim_{\epsilon\rightarrow 0} \Biggl[
\frac{1}{\pi}\frac{\partial_u U(u+\epsilon/2) \partial_u U(u-\epsilon/2)}{(U(u+\epsilon/2)-U(u-\epsilon/2))^2}  -\frac{1}{\pi}\frac{1}{\epsilon^2} \Biggr] .
\label{boundary-transformation}
\end{align} 
Here we have denoted the terms in the first parenthesis as $\langle:\mathcal{O}^{(U)}\mathcal{O}^{(U)}(U)  :\rangle$, since it is a regularized operator with respect to $U$ coordinate.
In addition we can show that the last term in (\ref{boundary-transformation}) becomes
\begin{align} 
\lim_{\epsilon\rightarrow 0} \Biggl[
\frac{1}{\pi}\frac{\partial_u U(u+\epsilon/2) \partial_u U(u-\epsilon/2)}{(U(u+\epsilon/2)-U(u-\epsilon/2))^2}  -\frac{1}{\pi}\frac{1}{\epsilon^2} \Biggr]
=\{U,u\}/6\pi.
\nonumber
\end{align} 
Thus $\langle :\mathcal{O}\mathcal{O}(u):\rangle$ has an anomalous transformation property equivalent to that of the bulk energy-momentum tensor (\ref{transformation}) and we can obtain an equation similar to (\ref{transformation-energy-momentum}) in case $U(u)=-e^{-\kappa u}$.

Now we consider a coordinate transformation $\tilde{V}=-1/V$, then the metric (\ref{2dbh}) becomes
\begin{align} 
ds^2=-\frac{4dU d \tilde{V}}{(\tilde{V}-U)^2}.
\end{align} 
This metric is the Poincar\'{e} coordinate in the light-cone gauge and the boundary is given by $\tilde{V}=U$.
Then we know that a point $\tilde{V}=U=0$ in the boundary is not a special point. 
Therefore $\langle:\mathcal{O}^{(U)}\mathcal{O}^{(U)}(U)  :\rangle$ should be regular at this point\footnote{Our regularity condition corresponds to a global AdS vacuum condition in \cite{Spradlin:1999bn}.} and this regularity condition requires
\begin{align} 
\langle T(u) \rangle=\frac{1}{4}\langle:\mathcal{OO}(u):\rangle
=
-\{U,u\}/24\pi =\frac{\kappa^2}{48\pi}.
 \label{flux-boundary}
\end{align} 
This is the boundary derivation of the energy flux and this result is consistent with the previous result (\ref{energy-flux}).
We can conclude that the Hawking radiation is derived from the anomalous transformation (\ref{boundary-transformation}) of the boundary operator and the regularity.\\

Lastly we comment about the locality of the relation (\ref{energy-momentum-boundary}).
The equation (\ref{correlation}) relates a local correlation function in the bulk with a non-local operator in the boundary. 
On the other hand, the equation (\ref{energy-momentum-boundary}) is a correspondence between a bulk local operator and a boundary local operator.
This is because the bulk operator is holomorphic and depends on $u$ only.
On the other hand, we can derive a similar correspondence for anti-holomorphic operator $\langle T(v) \rangle = \frac{1}{4}\langle :\mathcal{O}\mathcal{O}(v): \rangle$ and, in order to obtain all components of the bulk energy-momentum tensor at $(t,r_*)$, we have to know the information of the boundary operators at $t \pm r_*$ (see figure 1.).
Thus the local information of the bulk still corresponds to the non-local information of the boundary.\\
\begin{figure}
\begin{center}
\includegraphics{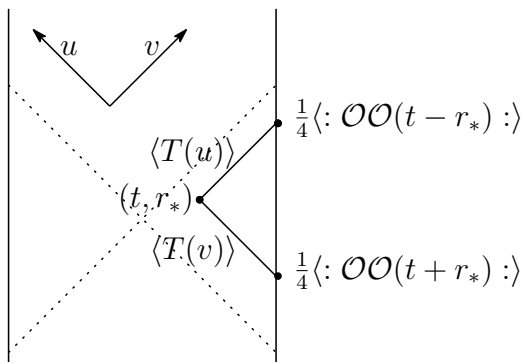}
\caption{The correspondence between the bulk operators at $(t,r_*)$ and the boundary operators at $(t \pm r_*)$. 
Note that $r_*$ is always negative.}
\end{center}
\end{figure}

\section{Boundary description of bulk higher-spin operators}
In this section, we discuss the construction of higher-spin operators in the bulk from boundary operators.
We can define holomorphic higher-spin operators which generate $W_\infty$ algebra as
\begin{align} 
J_{m,n}(u)=:\partial_u^m \phi(u,v) \partial_u^n \phi(u,v):.
\label{higher-spin-current}
\end{align} 
These currents satisfy anomalous conformal transformations \cite{Iso:2007kt} and these transformations are related to generalized trace anomalies \cite{Iso:2007nc,Bonora:2008nk,Bonora:2008he}.
As in (\ref{energy-momentum-boundary}), we can construct corresponding boundary composite operators,
\begin{align} 
\langle J_{m,n}(u) \rangle=\frac{1}{4}\langle: (-)^{m+n} \partial^{m-1}\mathcal{O} \partial^{n-1} \mathcal{O}(u):\rangle.
\end{align} 
It is easy to show that these boundary operators satisfy the same conformal transformations to the corresponding bulk operators as in (\ref{boundary-transformation}).
The relation between the anomalous transformations for the higher-spin currents and the full spectrum of thermal Hawking radiation has been shown in \cite{Iso:2007kt, Iso:2007hd, Iso:2007nf}.
Thus we can reproduce it from the boundary operators also.

\section{Conclusions and Discussions \label{conclusion}}
\setcounter{equation}{0}
In this letter, we found the correspondence between bulk and boundary composite operators.
If bulk operators are holomorphic or anti-holomorphic, they correspond to local operators in the boundary and if they are not holomorphic, they correspond to non-local operators in general.
Such correspondence will be important in the study of AdS$_2$/CFT$_1$.

By employing this correspondence, we evaluated the expectation value of the bulk energy-momentum tensor in terms of the boundary operator and reproduced the energy flux through Hawking radiation.
The anomalous transformation of the boundary composite operator (\ref{boundary-transformation}) is related to the energy flux.
This understanding will be helpful for the study of the black hole thermodynamics through the boundary CFT.
However, the meaning of the anomalous transformation in the boundary CFT$_1$ is not clear, since we have not used an explicit construction of the CFT$_1$.
The anomalous transformation of the energy-momentum tensor (\ref{transformation}) is derived from the trace anomaly ${T^\mu}_\mu=cR/24\pi$ in the bulk and it will be interesting to find a corresponding physics in the boundary.

It will also be interesting to investigate applications of our study to other instances of the AdS/CFT correspondence, especially AdS$_5$/CFT$_4$ and  AdS$_3$/CFT$_2$.
In these correspondences, we know the corresponding CFT better than CFT$_1$ and it is possible to construct concrete corresponding composite operators. 
Besides we can compare the point splitting regularization in this letter with the previous regularization in the AdS/CFT \cite{Chalmers:1999gc}.\\

We have discussed the out-going energy flux.
Now we mention the in-going flux which corresponds to $\langle T(v) \rangle$.
We can show that $\langle T(v) \rangle= \frac{1}{4}\langle : \mathcal{OO}(v) : \rangle$ as in (\ref{boundary-transformation}).
Note that $\frac{1}{4}\langle :\mathcal{OO}(u): \rangle$ is given by $\kappa^2/48\pi$ and it implies that $\frac{1}{4}\langle :\mathcal{OO}(v): \rangle$ is also $\kappa^2/48\pi$.
Thus the in-going energy flux from the outside of the black hole balances the out-going flux from the black hole and the total energy flux becomes $0$.
This result means that the vacuum we have selected is the Hartle-Hawking vacuum \cite{Hartle:1976tp, Maldacena:2001kr}.
In black hole physics, we can also consider the Unruh vacuum \cite{Unruh:1976db} in which the in-going flux vanishes at infinity.
It will be interesting to investigate the construction of the Unruh vacuum in the AdS/CFT correspondence.\\

Recently a new interpretation of Hawking radiation was proposed by Robinson and Wilczek \cite{Robinson:2005pd, Iso:2006wa, Iso:2006ut, Banerjee:2007qs, Iso:2008sq}.
They divided the region outside the horizon into two regions: near horizon region and the other region, and omitted the in-going modes  in the near horizon region.
Then the gravitational anomaly appeared there and they derived the energy flux as in the trace anomaly method.
By applying this prescription to our study, it may be possible to understand this method in terms of boundary theory.
However, the reason for omitting the in-going mode in the near horizon region is not clear in boundary description.
In \cite{Hamilton:2005ju,Hamilton:2006az,Hamilton:2006fh}, the relations between bulk correlation functions inside the horizon and boundary correlation functions were studied. 
Thus by considering these studies, it will be interesting to investigate whether the gravitational anomaly near the horizon actually appears or not.
\\

Lastly, we discuss the relation between the covariant energy-momentum tensor and the holomorphic energy-momentum tensor which we have studied in this letter.
In (\ref{energy-momentum}), we regularized the energy-momentum tensor such that it satisfies the holomorphy.
As a result, the regularization breaks covariance under conformal transformations and the anomalous term (the Schwarzian derivative) appears in (\ref{transformation}).
On the other hand, it is possible to regularize the energy-momentum tensor such that it satisfies the covariance \cite{Iso:2007nc}.
We denote the covariant energy-momentum tensor as $T_{uu}(u,v)$ and the relation between these two energy-momentum tensors is given by,
\begin{align}
 T_{uu}(u,v) =T(u) +\frac{c}{24\pi}
  \left(\partial_u^2 \varphi(u,v) - \frac{1}{2}(\partial_u \varphi(u,v))^2\right)
  \label{covariant}
\end{align}
Here we have rewritten the metric as $ds^2=-e^{\varphi(u,v)} dudv$.
This relation implies that the covariant energy-momentum tensor depends on the value of the background metric.
Thus in order to describe this covariant operator in terms of boundary operators, we have to express the metric by boundary operators and it will be difficult in general.

Note that if we evaluate the vacuum expectation value of the covariant energy-momentum tensor (\ref{covariant}) in the coordinate (\ref{2dbh}), we can show that it becomes $0$.
This is one feature of radiation in Rindler space.

\paragraph{Acknowledgements }
I would like to acknowledge useful discussions with Satoshi Iso, Shiraz Minwalla, Hiroshi Umetsu and Spenta Wadia.
I would also especially like to thank Gautam Mandal for useful discussions and for several detailed comments on the manuscript.

\appendix

\section{Derivation of smearing function}
\setcounter{equation}{0}
\subsection{Correlation function correspondence in AdS/CFT}
\label{appendix correlation}
In this appendix, we briefly summarize some properties of the Lorentzian AdS/CFT duality \cite{Balasubramanian:1998sn, Balasubramanian:1998de, Banks:1998dd} and correspondence between bulk and boundary correlation function.

Free massless scalar fields in AdS space have two solutions: a normalizable mode and a non-normalizable mode.
The non-normalizable mode has a non-zero classical value at the boundary and it acts as a source which deforms the CFT.
Thus we do not consider the non-normalizable mode in our study.
The normalizable mode of the free massless scalar field behaves as
\begin{align} 
\phi(z,x) \rightarrow z^{\Delta } \phi_0(x)
\end{align} 
near the boundary of AdS$_{d+1}$.
Here $z$ is a radial coordinate which vanishes at the boundary and $x$ is a $d$ dimensional coordinates.
This behavior shows that this field corresponds to a boundary operator $\mathcal{O}$ which has conformal dimension $\Delta=d$ \cite{Gubser:1998bc,Witten:1998qj}
\begin{align} 
\phi_0(x) \leftrightarrow \mathcal{O}(x).
\label{dual-operator}
\end{align} 
Then we have a correspondence between the local field in the bulk and a non-local operator in the boundary by employing a smearing function $K(x'|z,x)$,
\begin{align} 
\phi(z,x) \leftrightarrow \int dx' K(x'|z,x) \mathcal{O}(x').
\label{smearing }
\end{align} 
Note that $K(x'|z,x)$ is different from the bulk-boundary propagator for the non-normalizable mode in  \cite{Witten:1998qj}.
By using this relation, we can evaluate the bulk correlation function in terms of the corresponding boundary correlator \cite{Banks:1998dd,Balasubramanian:1999ri,Hamilton:2005ju},
\begin{align} 
\langle \phi(z_1,x_1) \phi(z_2,x_2) \rangle=\int dx_1' \int dx_2'
K(x_1' | z_1,x_1)K(x_2' | z_2,x_2)\langle \mathcal{O}(x_1') \mathcal{O}(x_2') \rangle.
\end{align} 
By using this relation, we describe bulk composite operators in terms of corresponding boundary operators.

\subsection{Smearing function for two dimensional free massless scalar field}
\label{appendix smearing }

We show a derivation of the smearing function (\ref{smearing function}) for two dimensional free massless field $\phi$.
The equation of motion for $\phi$ in $(u,v)$ coordinate is given by
\begin{align} 
\partial_u \partial_v \phi=0.
\end{align} 
Then the solution is written as
\begin{align} 
\phi(u,v)=f_v(v)+f_u(u).
\end{align} 
Here $u = t-r_*$ and $v=t+r_*$, and $r_*$ corresponds to $z$  in the previous subsection.
$f_v$ and $f_u$ are arbitrary analytic functions which will be fixed by boundary conditions.
We can expand this solution with respect to $r_*$ near the boundary,
\begin{align} 
\phi(u,v) \rightarrow f_v(t)+f_u(t)+r_* \left( \partial f_v(t) - \partial f_u(t) \right)+O(r_*^2).
\end{align} 
On the other hand, the normalizable mode for this field satisfies the boundary condition
\begin{align} 
\phi(u,v) \rightarrow r_* \phi_0(t).
\end{align} 
It implies $f_v=-f_u$ and $\partial f_v=\phi_0/2 $.
Then the solution can be described as
\begin{align} 
\phi(u,v) = -\frac{1}{2}\int_{v}^{u} dt'  \phi_0(t') =\int_{-\infty}^{\infty}  dt'   K(t'|u,v)\phi_0(t'),
\end{align} 
and we obtain the smearing function (\ref{smearing }) for this solution as 
\begin{align} 
K(t'|u,v)=\frac{1}{2}\left( \theta(t'-u)-\theta(t'-v) \right) ,
\end{align} 
Here $\theta(t)$ is the step function,
\begin{align} 
\theta(t)= \left\{ \begin{array}{ll}
1 & (t\ge 0) \\
0 & (t<0) \\
\end{array} \right.
.
\end{align}


\bibliographystyle{JHEP}
\bibliography{anomaly}

\end{document}